\documentclass[]{article}

\usepackage{graphicx} 

\begin{document}
\title{Towards the Heider balance -- a cellular automaton with a global neighborhood}

\author{Malgorzata J. Krawczyk, Krzysztof Kulakowski, Zdzislaw Burda\\[.5cm]
AGH~University~of~Science~and~Technology,\\Faculty~of~Physics~and~Applied~Computer~Science,\\ al.~Mickiewicza~30,~30-059 Kraków,~Poland.
}
%
%
%
%
\date{}


\maketitle

\begin{abstract}
We study a simple deterministic map that leads a fully connected network to Heider balance. The map is realized by an algorithm that updates all links synchronously in a way depending on the state of the entire network. We observe that the probability of reaching a balanced state increases with the system size $N$. Jammed states become less frequent for larger $N$.  The algorithm generates also limit cycles, mostly of length $2$, but also of length $3$, $4$, $6$, $12$ or $14$. We give a simple argument to estimate the mean size of basins of attraction of balanced states and discuss symmetries of the system  including the automorphism group as well as gauge invariance of triad configurations. We argue that both the symmetries play an essential role in the occurrence of cycles observed in the synchronous dynamics realized by the algorithm.
\end{abstract}

\section{Introduction}

In recent years social phenomena have become an attractive field of research for scholars in statistical physics and computer science \cite{1,2,3,4}. This interest is motivated by the spreading awareness that most important problems faced by  mankind are created by people themselves. On the other hand, the emerging science of complexity \cite{5} widens the list of algorithms and methods applicable to dynamic processes in social networks \cite{1,6}.

\medskip

Here we focus on modeling the process of removal a cognitive dissonance \cite{7} in social relations within a fully connected social network. During this process, the network evolves towards the so-called Heider balance called also structural balance, where the dissonance is removed \cite{8,9,10}. To the best of our knowledge, there are five different algorithms designed to model this process: two Monte-Carlo algorithms \cite{11,12} and three deterministic ones \cite{13,14,15,16}.  The first two distinguish only positive (friendly) and negative (hostile) relations. The next two are deterministic and rely on numerical solutions of differential equations for relations being represented by real numbers, that are either limited to the range $(-1,+1)$ \cite{13} or may tend to $\pm$ infinity in a finite time \cite{14}. The advantage of the latter approach \cite{14} is that the equations may be solved analytically. The last method \cite{15,16} is a deterministic synchronous cellular automaton, where the states of links are updated based on a local rule. Recently the Monte-Carlo simulations \cite{12} have been generalized by including thermal noise \cite{17,18} and a specific aging of the relations \cite{19}.  On the other hand, the differential equations \cite{13} have been used to simulate asymmetric links, which mimic unreciprocated relations \cite{20,21}.

\medskip

In a balanced state, the relations in each triad of actors conform to the following rules \cite{22}:
\begin{itemize}  
\item A friend of my friend is my friend,
\item An enemy of my friend is my enemy,
\item A friend of my enemy is my enemy,
\item An enemy of my enemy is my friend.
\end{itemize}
When translated to signs of relations, the four rules allow to distinguish balanced from imbalanced triads. In the balanced state, the product of the three signs is +1 in each triad. The process of reducing the dissonance in the network is equivalent to a minimization of the negative sum over all triads
\begin{equation}
U=-\sum_{i<j<k}^N \Delta_{ijk}
= - \sum_{i<j<k}^N s_{ij}s_{jk}s_{ki}
\label{sum}
\end{equation}
where the sign $s_{ij}=\pm 1$ represents the symmetric relation between actors located in the nodes $i$ and $j$ of the network. One can think of $U$ as energy of the system.  As shown in \cite{9}, when all triads are balanced then the whole network is split into two groups, internally friendly and mutually hostile. Clearly, in such states there is no cognitive dissonance about who is friend and who is enemy. 

In this paper we study a discrete time evolution, which is deterministic and synchronous like for cellular automata \cite{15},  however -- as in \cite{11,13} -- the basic update rule depends on the entire state of the network
\begin{equation}
s_{ij}(t+1)=\left\{
\begin{array}{ll}
{\rm sgn} \left( \sum_{k\ne i,j}^N s_{ik}(t) s_{kj}(t) \right) & {\rm if} \quad  \sum_{k\ne i,j}^N s_{ik}(t) s_{kj}(t) \ne 0 \\  
 s_{ij}(t)  & {\rm otherwise}  
\end{array}
\right.
\label{e1}
\end{equation}
for all pairs $i\ne j$.  The sum runs through all vertices $k$ other than $i$ and $j$. The most important difference to a typical cellular automaton is that neighborhood of an updated edge (\ref{e1}) increases with the system size. Hence the term 'global' in the title. The update rule (\ref{e1}) is equivalent to
 \begin{equation}
s_{ij}(t+1)= \left\{ 
\begin{array}{rl} 
s_{ij}(t)  & {\rm if}  \quad  - U_{ij}(t)  \equiv \sum_{k\ne i,j}^N \Delta_{ijk}(t) \ge 0 \\
-s_{ij}(t) & {\rm otherwise}  
\end{array}
\right.
\label{e2}
\end{equation}
where the sum in the last equation runs over all triads sharing the edge $ij$.
The update rule (\ref{e1}), or equivalently (\ref{e2}), can be interpreted as a majority rule according to which the sign of the edge $ij$ adjusts to the sign of majority of triads it is a part of. Clearly, if the updates (\ref{e2}) were applied asynchronously, edge by edge, they would either leave energy (\ref{sum}) unchanged $U(t+1)=U(t)$ if $U_{ij}(t)\le 0$, or reduce it to
\begin{equation}
U(t+1) = U(t) - 2U_{ij}(t),
\label{UUU}
\end{equation}
if $U_{ij}(t)>0$, as a result of changing the signs of triads containing the edge $ij$. The update rule (\ref{e2}) would never increase energy.  
For synchronous updates this is not the case. The synchronous dynamics is much more complex which makes it very interesting.  

Let us  also note that for odd $N$ the second line in Eq. (\ref{e1}) may be omitted because the sum of odd number of terms $\pm 1$ is never zero. For even $N$ the second line of Eq. (\ref{e1}) means ''do nothing if there is no majority''.  This choice is similar to the Nash equilibrium where an actor does not change strategy if s/he cannot profit from the change \cite{str}.  This choice is also akin to the delay of updating, applied in \cite{23} to eliminate 'unstable' attractors in Boolean networks.  We shall mention an alternative solution later while discussing symmetries of the system.

The rationale behind Eq. \ref{e1} is the same as in \cite{11,12,13,14,15,16}. Namely, the relation $s_{ij}$ of $i$ with $j$ is improved, if the relations $s_{ik}$ and $s_{kj}$ are both friendly ("friend of my friend") or both hostile ("enemy of my enemy").  In two other cases ("friend of my enemy" or "enemy of my friend") the relation $s_{ij}$ is deteriorated. Here, as in \cite{13}, this consultation of relation $s_{ij}$ is extended over all agents $k$ of the network. The difference with \cite{13} is that here links are updated in one time step. The same kind of updating  was applied in \cite{15} for a local neighborhood.  Our approach is complementary to the previous ones \cite{11,12,13,14,15,16}.  As we demonstrate below the results obtained within this approach are qualitatively new. They broaden the spectrum of the types of behavior that can be observed in evolution of social networks that are trying to achieve Heider balance.

The rest of the paper is composed as follows. In the next section we graphically demonstrate that for $N=3,4$ the system evolves immediately (in one time step) towards a balanced state, while for $N=5$, short limit cycles,  of length $2$, are also possible.  In Section 3 we discuss symmetries of the system.  In Section 4 we report the results on the influence of the system size on the number of transient iterations and on the probability of balanced states. Section 5 is devoted to the limit cycles observed in larger systems.  The paper is concluded by a summary and a discussion of results in Section 6.

\section{Small systems}

\begin{figure}
\begin{center}
\includegraphics[width=0.6\columnwidth]{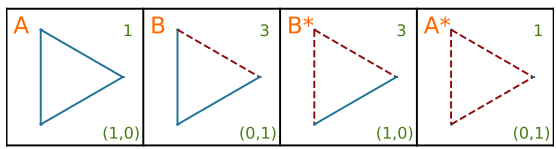}\\[.51cm]
\includegraphics[width=0.4\columnwidth]{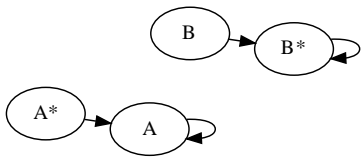} 
\caption{Top: Configuration classes for $N=3$. The class cardinalities are given in the upper right corners of the plaquettes. In the lower right corner, in brackets, the numbers are given of balanced and imbalanced triads. The same coding is applied below. A starred class $X^*$ is obtained from $X$ by swapping all signs: plus one to minus one and minus one to plus one. Bottom: Flow diagram representing transitions between the classes, under the synchronous update rule (\ref{e1}). }
\label{N=3}
\end{center}
\end{figure}

For $N=3$, there are eight distinct configurations: $(+++)$, $(++-)$, $(+-+)$, $(-++)$, $(--+)$, $(-+-)$, $(+--)$ and $(---)$ that can be grouped into four distinct classes of indistinguishable unlabelled configurations, that we denote by $A,A^*,B,B^*$, see Fig. \ref{N=3}. The notation with the star superscript will become clear later. The class $A$ has three positive links, $B$ has two, $B^*$ has one, and $A^*$ has no positive links.  The classes $A,B^*$ are balanced while $A^*$ and $B$ are imbalanced.  The class cardinality is the number of distinct labelled configurations.  The cardinalities of the classes $A$ and $A^*$ are equal to one while of  $B$ and $B^*$ - to three.  The transitions between classes generated by the synchronous majority rule (\ref{e1}) are shown in Fig. \ref{N=3}. 
One can also list all configurations for $N=4,5,\ldots $. Let us first discuss in detail the case $N=5$, which illustrates a variety of interesting effects. Later we will return to $N=4$. The distinct unlabelled classes on the complete graph $K_5$ and collected in Table~\ref{N=5}.  In the table one can also see the multiplicities of the classes as well as the corresponding numbers of positive and negative triads, which are displayed in parentheses $(\Delta_+,\Delta_-)$.  It is not a complete list of configurations for $N=5$, because we have left out the configurations that can be obtained from those shown in the table, swapping the positive and negative links. Such configurations will be denoted by star; for instance $A^*$ is obtained from $A$ so it consists only of negative edges. $B^*$ is obtained from $B$ and thus has a single positive edge and nine negative ones, etc.. We shall refer to starred classes as dual classes. A change of signs of all edges applied twice leaves them unchanged, then we have $A=A^{**}$. The complete graph $K_5$ has $10$ links, and thus there are $2^{10} = 1024$ configurations.  Adding cardinalities of all classes from Table~\ref{N=5} and of the corresponding dual classes  one finds that the sum is indeed equal to $1024$.  One should note that the  classes $O$ and $Q$ are self-dual, that is $O=O^*$ and $Q=Q^*$, and that  the classes $P$ and $T$, as well as $R$ and $S$ are mutually dual $P=T^*$ and $R=S^*$. Therefore, in order to avoid double counting, the multiplicities of these classes should be counted once, as opposed to those of other classes from the table, which should be counted twice in order to take into account contributions from the class and its dual partner.

\begin{figure}
\begin{center}
\includegraphics[width=0.80\columnwidth]{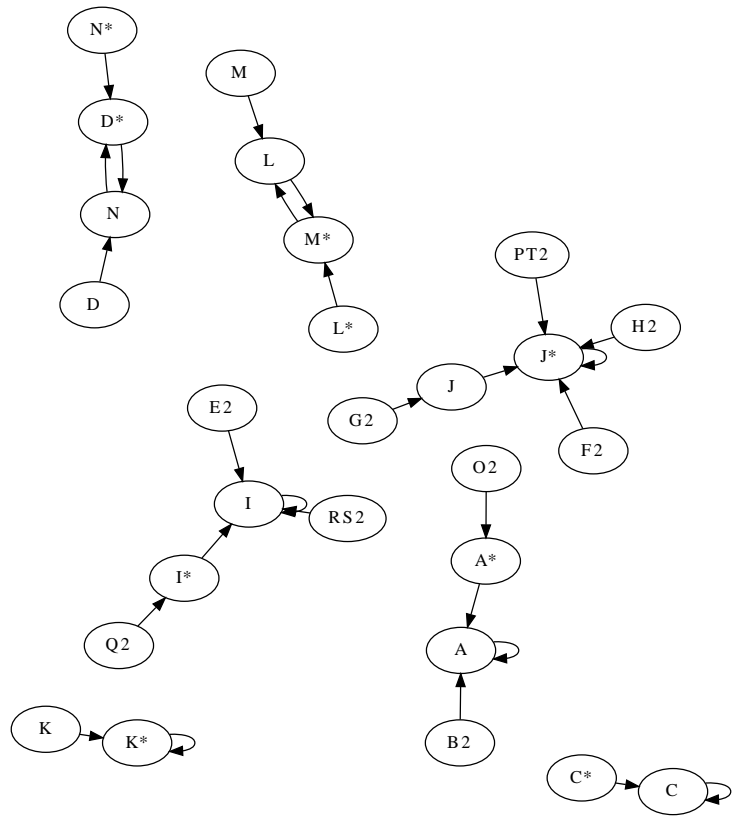} 
\caption{Flow diagram representing transitions between unlabelled classes for $N=5$, under the synchronous update rule (\ref{e1}).}
\label{baseny}
\end{center}
\end{figure}

Let us now study the action of the transformation rule (\ref{e1}) on the classes. The result is illustrated in Fig. \ref{baseny} as a flow diagram which indicates which class is mapped into which. The diagram has seven connected components, each representing a different basin of attraction. The symbol $F2$  means a union of two classes, $F2=F \cup F^*$, and similarly for other letters. The double letters $PT2$ and $RS2$ are used for dual classes to underline 
that $T2=P2$, and similarly $R2=S2$.
\begin{table}
\begin{center}
\includegraphics[width=\columnwidth]{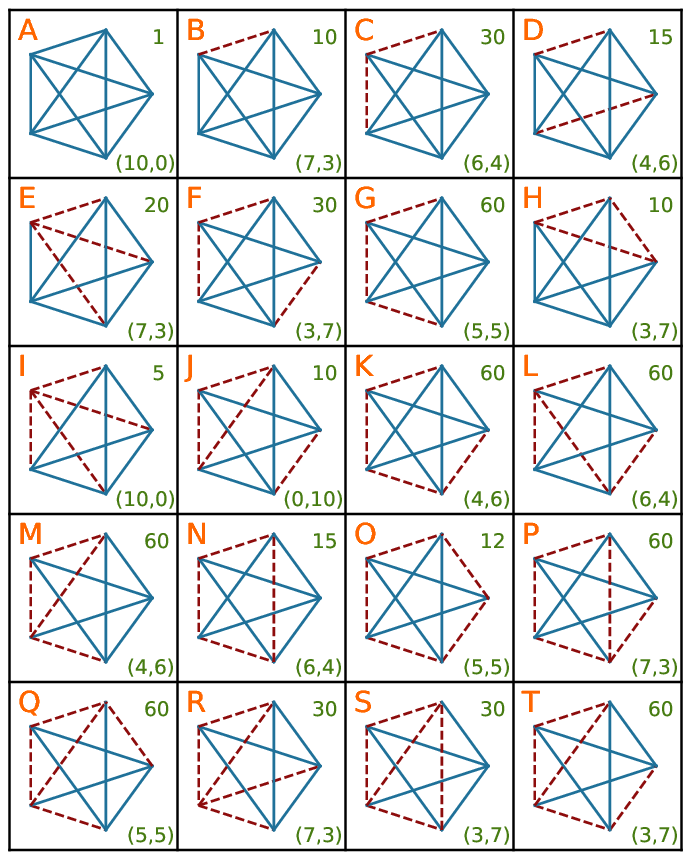} 
\caption{Classes of unlabelled configurations for $N=5$, their cardinalities and the numbers of balanced and imbalanced triads. }
\label{N=5}
\end{center}
\end{table}

\begin{figure}
\begin{center}
\includegraphics[width=0.45\columnwidth]{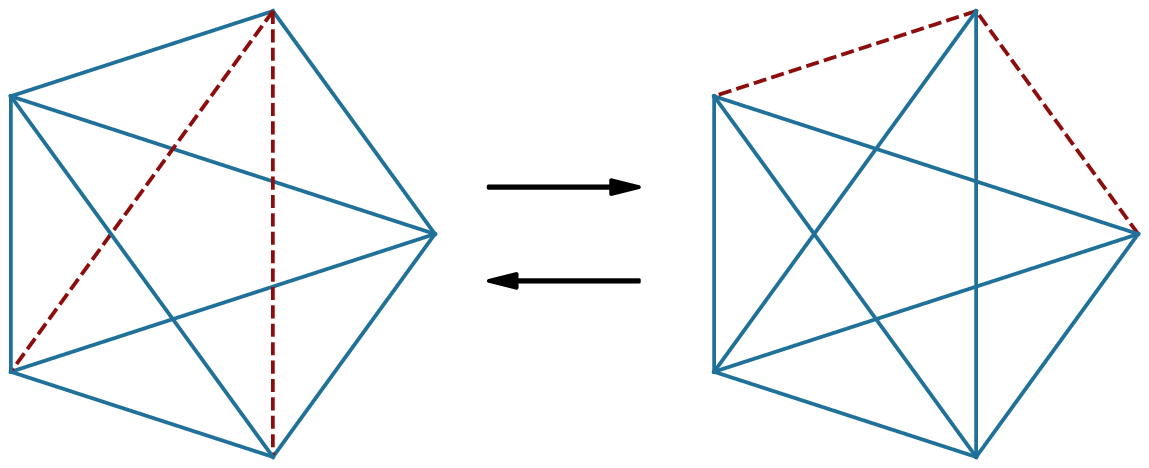}  
\quad
\includegraphics[width=0.45\columnwidth]{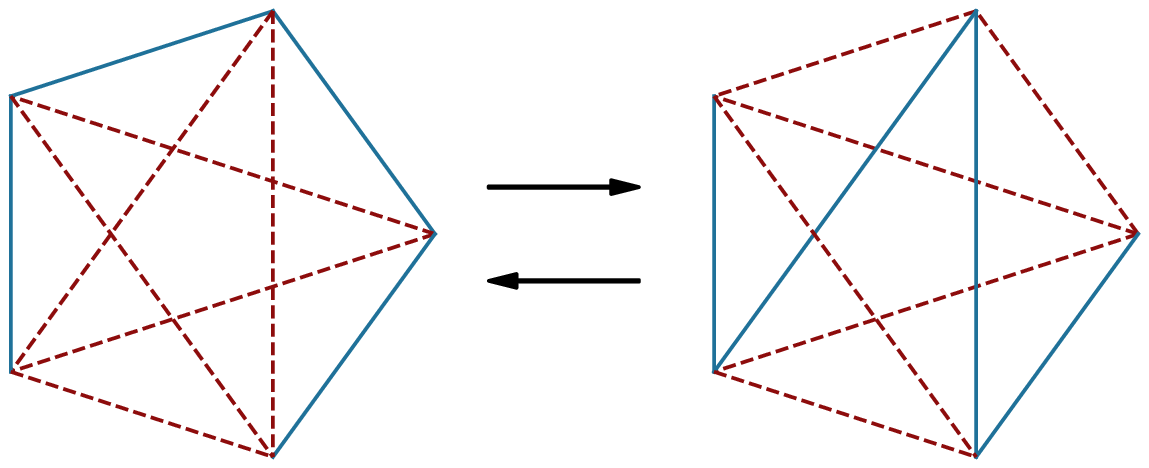} 
\caption{Internal limit cycles within $C$ class (left) and $K^*$ class (right). }
\label{tau2}
\end{center}
\end{figure}

The attractors in Fig. \ref{baseny} can be divided into two different types: absorbing classes, and limit cycles of order two. There are five absorbing classes: $A$, $I$, $J^*$, $C$ and $K^*$ and two limit cycles of order two ($D^*$,$N$), and ($L$,$M^*$).  Three absorbing classes are balanced: $A$,$I$ and $J^*$ and two are imbalanced $K^*$, $C$. Interestingly the absorbing classes $K^*$ and $C$ correspond in fact to the limit cycles of length two which are internal cycles between different configurations from the same class.  The cycles are shown in Fig. \ref{tau2}.  To summarize, for $N=5$ we observe absorbing states, as well as limit cycles of length two. 

\section{Symmetries}
The discussed system has a high symmetry. The basic symmetry is related to the automorphism group of the complete graph which forms the skeleton of the system.  The other symmetry is related to gauge invariance of triads. Let $\{ \Delta_{ijk}\}_{i<j<k}$ be the list of values of the products $\Delta_{ijk} = s_{ij} s_{jk} s_{ki}$ for all triads. The following transformation applied to all edges $ij$
\begin{equation}
s_{ij} \rightarrow s'_{ij} = \sigma_i s_{ij} \sigma_j  
\label{gauge}
\end{equation}
where $\sigma_i = \pm 1$, for $i,j=1,\ldots,N$, 
leaves all the products invariant
\begin{equation}
\Delta'_{ijk} =s'_{ij} s'_{jk} s'_{ki} = 
 s_{ij} s_{jk} s_{ki} = \Delta_{ijk} 
\end{equation} 
because each sigma is squared yielding $\sigma_i^2=\sigma_j^2=\sigma_k^2=1$.
As a consequence the same set of  values
\begin{equation}
\{ \Delta_{ijk}\}_{i<j<k} \rightarrow \{ \Delta'_{ijk}\}_{i<j<k}= \{ \Delta_{ijk}\}_{i<j<k} 
\end{equation}
can be represented by distinct edge configurations. Borrowing terminology from high energy physics the configurations representing the same set of $\Delta$'s can be called Gribov copies \cite{gri}.  Since one can have two values $\sigma_i=\pm 1$ (\ref{gauge}) for $(N-1)$ vertices, there are $2^{N-1}$ Gribov copies.  For one vertex, does not matter which one, one should have only one value $\sigma_i=1$, because as one can see from Eq. \ref{gauge}, when all $\sigma$'s are equal to minus one, all edges acquire the same values, $s'_{ij}=s_{ij}$, as in the primary configuration.

To see how it works, let us consider again the case $N=5$. There are $2^4=16$ copies of each configuration. The configuration from class $A$ can be easily copied to a configuration from class $I$ by choosing $\sigma_i=-1$ for one vertex and $\sigma_j=1$ for all remaining vertices $j\ne i$. In this way all edges attached to $i$ will change sign yielding a configuration from the class $I$. If we choose $\sigma_i=\sigma_j=-1$ for two vertices, and $\sigma_k=1$ for all remaining ones $k\ne i,j$, we obtain a configuration from class $J^*$. There is one configuration in class $A$, five in class $I$ and ten in class $J^*$, altogether $16$ copies.  In a similar way one can identify copies of any (labelled) configuration. 

Coming back to the internal cycles in class $C$ and class $K^*$ that we discussed earlier (Fig. \ref{tau}), one can easily see that the consecutive configurations are Gribovs copies of one another.  Actually this is also a feature of many limit cycles that we observe for higher $N$.

\begin{figure}
\begin{center}
\includegraphics[width=0.45\columnwidth]{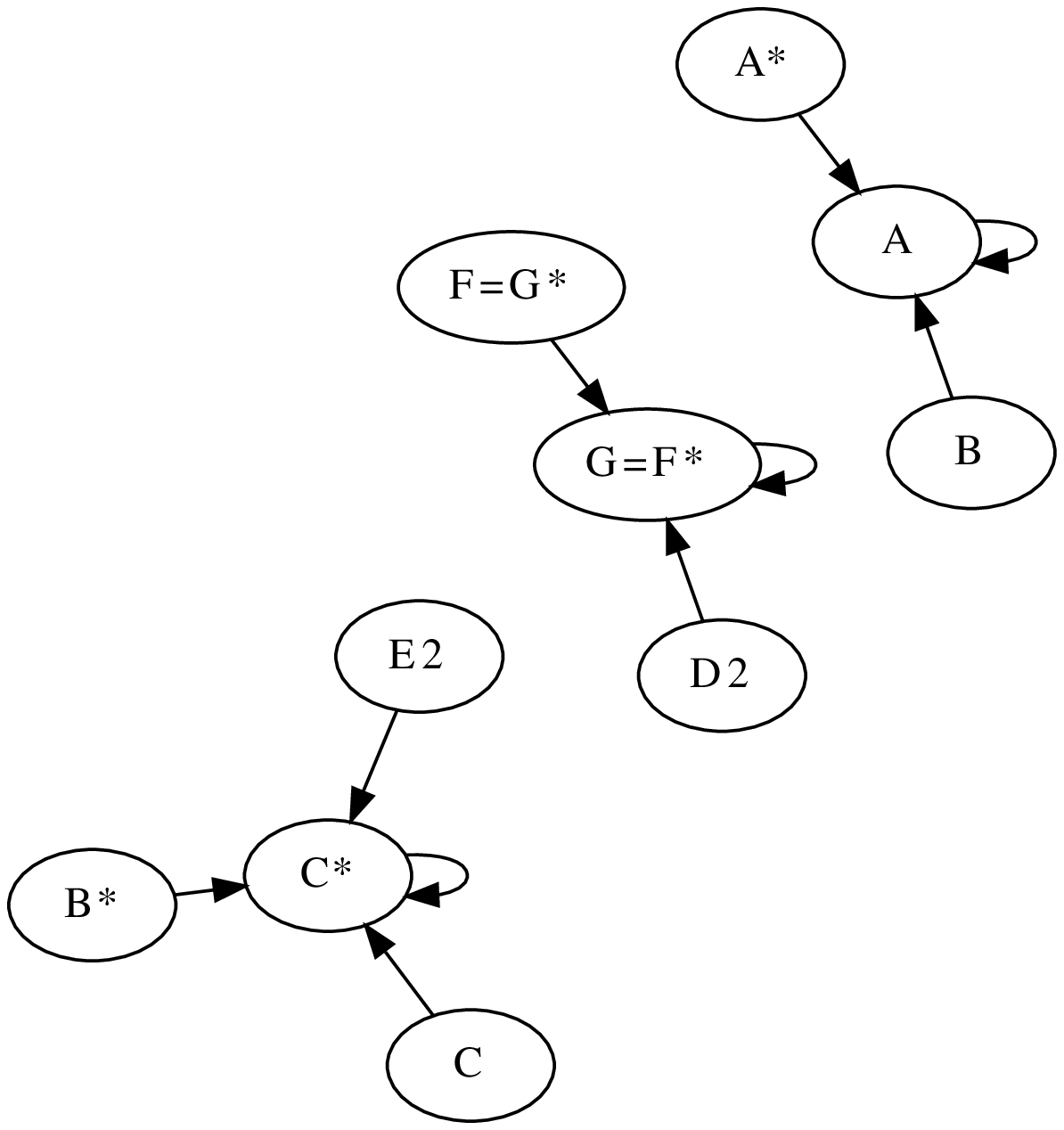}  \quad
\includegraphics[width=0.45\columnwidth]{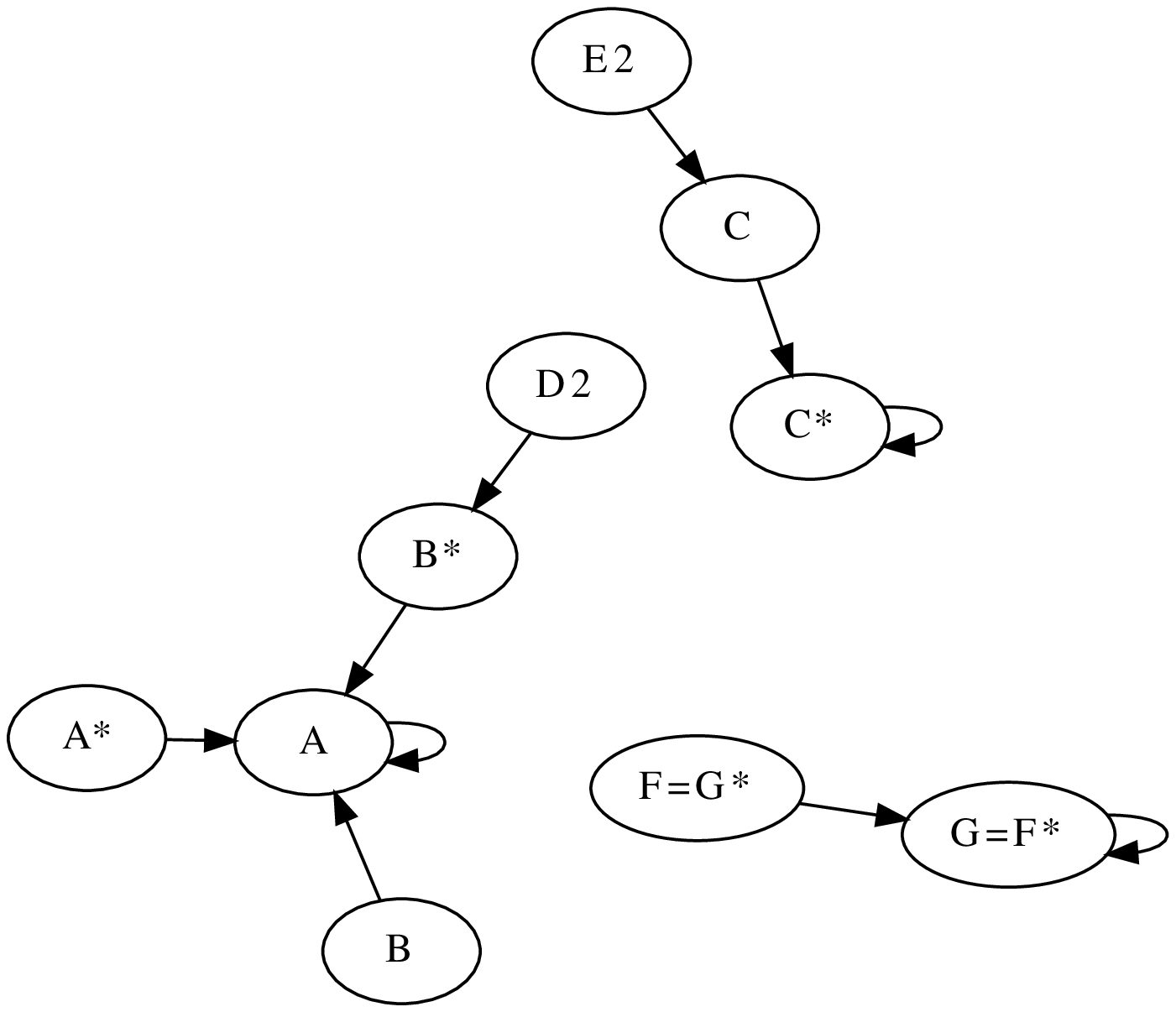} 
\caption{Left: Flow diagram for the update rule (\ref{e2}). Right: Flow diagram for
the update rule  (\ref{even}). }
\label{diff}
\end{center}
\end{figure}

We conclude this section by discussing  yet another issue related to the symmetry of the system. As mentioned, for odd $N$ the update rule (\ref{e1}) amounts to a pure majority rule
\begin{equation}
s_{ij}(t+1)= {\rm sgn} \left(\sum_{k\ne i,j}^N s_{ik}(t) s_{kj}(t)\right) \ .
\end{equation}
The results of the action of this rule on a configuration and a configuration obtained from it by changing globally all signs $s_{ij} \rightarrow s'_{ij} = -s_{ij}$, are identical, since the additional minus signs cancel on the right hand side of the equation.  This reduces the complexity of the problem.  The reduction does not apply for even $N$ since the second line of the update rule (\ref{e1}) depends on individual signs. As a result, it has to be considered separately how this rule works on the configuration and its double partner.  One can however slightly modify the update rule (\ref{e1}) in order to restore the symmetry with respect to the global change of all signs
\begin{equation}
s_{ij}(t+1)= {\rm sgn'}\left( \sum_{k\ne i,j}^N s_{ik}(t) s_{kj}(t)\right) \ .
\label{even}
\end{equation}
where the function ${\rm sgn}'$ is a modified sign function such that  ${\rm sgn}'(x) =1$ for $x\ge 0$ and ${\rm sgn}'(x)=0$ otherwise. For large $N$ the number of cases when the argument is exactly equal to zero, decreases, so one can expect that for large $N$ the difference between the effects of applying the modified rule (\ref{even}) and the original one (\ref{e1}) disappears. In Table \ref{N=4} we list configuration classes for $N=4$ and in Fig. \ref{diff} we illustrate the difference between the basins of attraction for the two rules. We see that some states move from one basin to another.
\begin{table}[h]
\begin{center}
\includegraphics[width=.6\columnwidth]{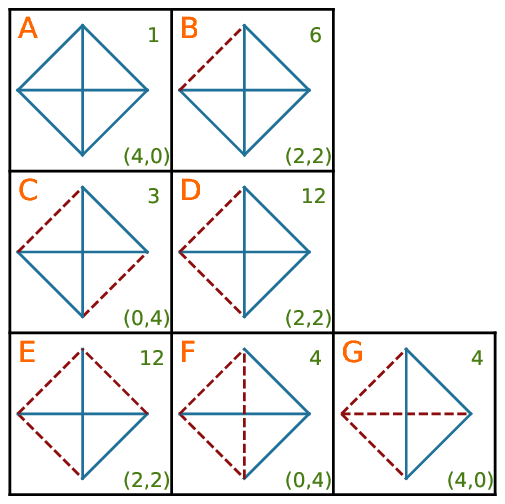} 
\caption{Unlabelled configuration classes for $N=4$.}
\label{N=4}
\end{center}
\end{table}

\section{Size effects}

\begin{figure}[!hptb]
\begin{center}
\includegraphics[width=\columnwidth]{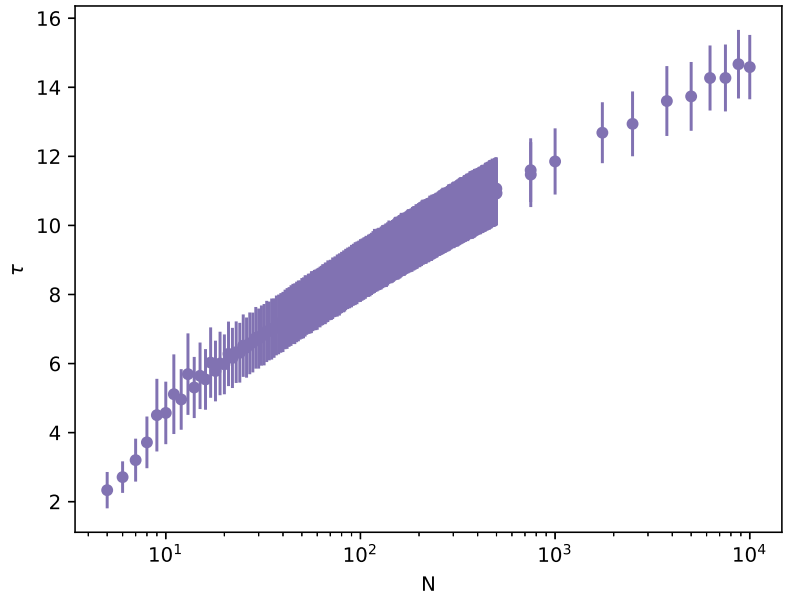} 
\caption{The average number, $\tau=\tau(N)$, of transient time steps versus the systems size, $N$. The mean $\tau(N)$ was calculated from $10^4$  process repetitions for $N<30$, $5\times 10^3$ -- for $30<N<10^3$, $400$ -- for larger $N$. }
\label{tau}
\end{center}
\end{figure}

\begin{figure}[!hptb]
\begin{center}
\includegraphics[width=\columnwidth]{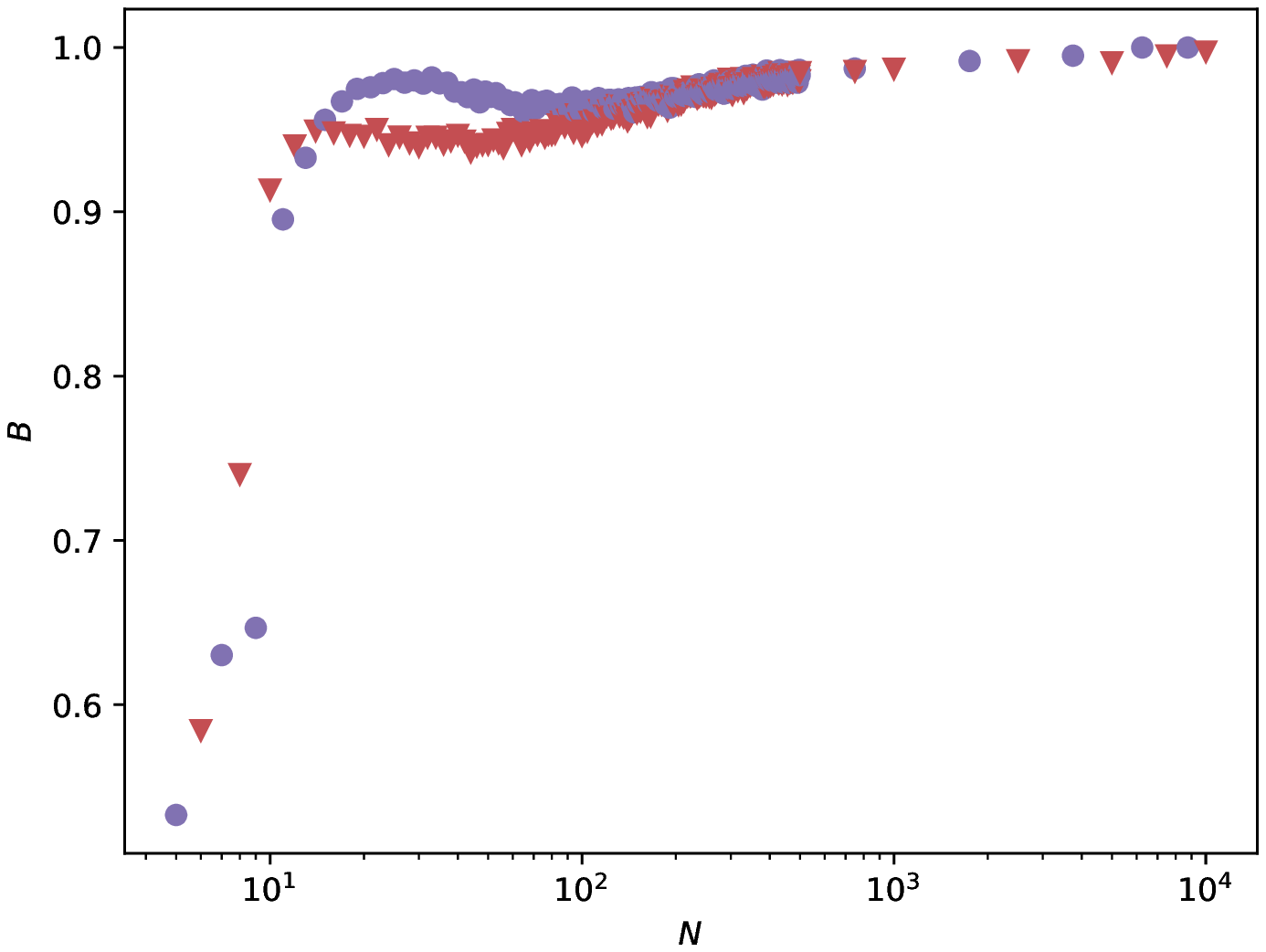} 
\caption{
The fraction, $B$, of balanced states versus the system size, $N$. The results for even $N$'s are plotted with circles, while for even $N$'s with triangles.  The statistics is the same as in Fig. \ref{tau}.}
\label{cc}
\end{center}
\end{figure}

The number of $L$ links is $N (N-1)/ 2$ and the update (\ref{e1}) of each link requires the calculation of a sum of $N-2$ terms (\ref {e1}), so the computational complexity of a single update of the whole system is $O(N^3)$.  Denote $\tau$ the number of transient updates till a fixed point or a limit cycle is reached.  Numerical results on average $\tau(N)$, for a system with $N$ nodes, are shown in Fig \ref{tau}. Let us make a rough estimate of the computational cost of transient part of the process.  $\tau(N)$ seems to increase logarithmically for large $N$, so the mean computational cost of reaching either a fixed point or a cycle is $O( N^3\log N)$.  This is more than $N^{4/3}$ for the Local Triad Dynamics, $p=1/2$ \cite{11} where $p$ is the probability of converting $(++-) \to(+++)$.  On the other hand, this  is much less than the cost of solving numerically $N^2/2$ 
equations \cite{13}.
We have also evaluated the fraction $B$ of balanced states among the steady states. The results are shown in Fig. \ref{cc}. In general, the fraction $B$ increases with $N$.  This result is similar to the result \cite{11} that the probability of imbalanced steady (jammed) states decreases with the system size. The values of $B$ follow different patterns for odd and even values of $N$.  The even-odd effect comes from the second line of Eq. \ref{e1}, which for even $N$ prevents some changes towards the balance.

\section{Limit cycles}

\begin{figure}
\begin{center}
\includegraphics[width=\columnwidth]{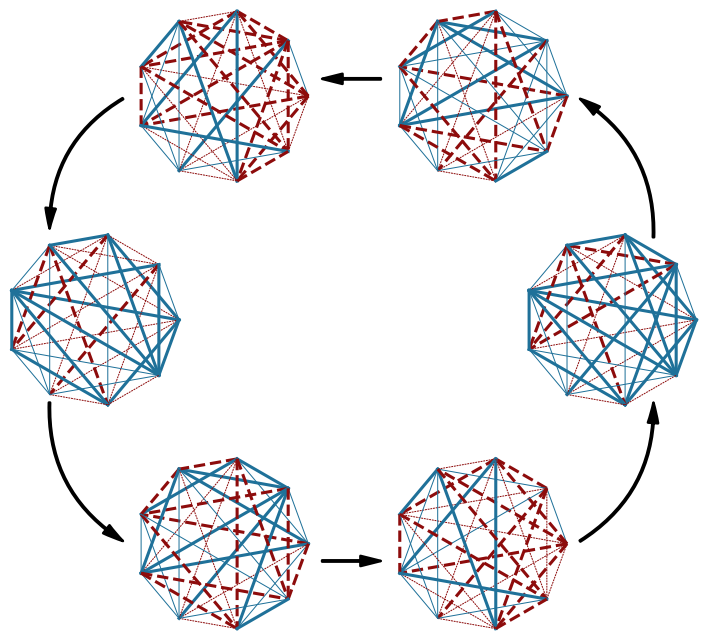} 
\caption{An example of a non-trivial limit cycle. In the figure we show a cycle of length $6$ for $N=9$.  The edges that have changed in the last step are drawn with a thick line, and those that have not changed - with a thin line.}
\label{ccc}
\end{center}
\end{figure}

Obviously a limit cycle cannot be balanced, because balanced states are absorbing, as follows from (\ref{e1}). For a Monte-Carlo evolution there are no limit cycles, as the system is driven either to a balance, or a jammed state \cite{11,12}.  Here we observe both jammed states and limit cycles. An example of a limit cycle is shown in Fig. \ref{ccc}.
In Table \ref{N=9}, detailed statistics of states leading to different attractors is presented. From these data, the following new observations can be deduced:

\noindent
-- the increase of the fraction, $B$, of balanced states with $N$ for $14<N<100$ is slow and it is visible only for larger $N$, as shown in Fig. \ref{cc}. In this smaller range of $N$, $B$ is slightly larger for odd values of $N$; 

\noindent
-- on the contrary, the fraction, $J$, of jammed states is larger for even values of $N$;

\noindent
-- the fraction $C$, of limit cycles decreases with $N$, so it is much easier to encounter a limit cycle for small systems.  The number of limit cycles seems to be slightly larger for odd values of $N$ than for
comparable even values.

For $N=5$ and larger, we observe limit cycles of length $L_c=2$. For $N=7-11$ also   limit cycles of length $3$ occur. Further, $L_c=4$ has been encountered for $N=9,10,11$ and $13$, $L_c=6$ for $N=9$ and $11$, and $L_c=12$ for $N=9$. The longest limit cycles we observed were of length $14$ for $N=11$ (Tab. \ref{t5}).

\begin{table}
\begin{tabular}{llll}
N & B & J & C \\
\hline
5 & 0.5329 & 0.0 & 0.4671 \\
6 & 0.5844 & 0.2385 & 0.1771 \\
7 & 0.6302 & 0.0 & 0.3698 \\
8 & 0.7403 & 0.0412 & 0.2185 \\
9 & 0.6468 & 0.0150 & 0.3382 \\
10 & 0.9135 & 0.0378 & 0.0487 \\
11 & 0.8954 & 0.0030 & 0.1016 \\
12 & 0.9403 & 0.0335 & 0.0262 \\
13 & 0.9329 & 0.0011 & 0.0660 \\
14 & 0.9494 & 0.0359 & 0.0147 \\
15 & 0.9560 & 0.0008 & 0.0432 \\
16 & 0.9482 & 0.0397 & 0.0121 \\
17 & 0.9672 & 0.0009 & 0.0319 \\
18 & 0.9468 & 0.0424 & 0.0108 \\
19 & 0.9748 & 0.0019 & 0.0233 \\
20 & 0.9462 & 0.0448 & 0.0090 \\
21 & 0.9758 & 0.0029 & 0.0213 \\
22 & 0.9504 & 0.0417 & 0.0079 \\
23 & 0.9783 & 0.0045 & 0.0172 \\
24 & 0.9410 & 0.0500 & 0.0090 \\
25 & 0.9808 & 0.0065 & 0.0127 \\
26 & 0.9457 & 0.0478 & 0.0065 \\
27 & 0.9786 & 0.0094 & 0.0120 \\
28 & 0.9421 & 0.0523 & 0.0056 \\
29 & 0.9800 & 0.0097 & 0.0103 \\
30 & 0.9398 & 0.0544 & 0.0058 \\
50 & 0.9412 & 0.0552 & 0.0036 \\
75 & 0.9670 & 0.0302 & 0.0028 \\
100 & 0.9466 & 0.0508 & 0.0026 \\
\hline
\end{tabular}
\caption{The fraction of balanced states $B$, of jammed states $J$, and of limit cycles $C$, as dependent on the system size $N$.
 The statistics is the same as in Fig. \ref{tau}.} 
\label{N=9}
\end{table}
To get more insight into the case $N=9$, we enumerated all initial configurations and calculated the numbers of initial conditions which lead to cycles of given length. All initial $2^{36}$ link configurations have been browsed. The results are given in Tab. \ref{t2}.  Clearly, the sums of numbers $c_k$ in each row are equal to $36!/N_1!/(36-N_1)!$, where $N_1$ is the number of positive links (and $N_-1 = L-N_1$ is the number of negative links).

The cycle distribution does not change when we interchange $N_1$ and $N_{-1}$. This is a demonstration of the global symmetry of Eq.\ref{e1} under the change of signs of all links. The symmetry holds only for odd values of $N$. For even values the second line of Eq.\ref{e1} breaks the symmetry, as discussed in Section 3.

We checked that for $N=9$ even for long cycles, the energy $U$ either remains constant during a cycle or takes at most two values. For $L_c=2$, these values which appear alternately are -144 and -36; the others are -312, -132, -120, -108 and -12.  For $L_c=3$ the energy remains constant during each cycle and its value is -84, -36 or +48. For $L_c=4$ a possible cycle of $U$ is -72, -72, +24, +24; the energy fixed points are -72, -60 or -24. For $L_c=6$ either $U=+12$, or a cycle appears: 0, 0, 0, +12, +12, +12. Finally $L_c=12$ we get only a fixed point $U=-36$. 

To complete the overall picture, we determined the cycles lengths also for $N=7,11$ with better statistics. For $N=7$ we browsed all initial configurations.  For $N=7$ the longest cycle has length $L_c=3$. This cycles of length $3$ occur in $368640$ out of $2^{21}$ cases (giving approximately frequency $0.176$).  For $N=11$ we sampled uniformly $10^{10}$ initial states out of all $2^{55} \approx 0.36 \cdot 10^{17}$ states, so the minimal frequency that can be detected is $O(10^{-7})$. The longest cycles recorded are of $L_c=14$.  The frequencies of different values of $L_c$ are given in Table \ref{t5}. While analysing the table one has to remember that some objects which are less frequent than the limiting frequency $O(10^{-7})$, may not show up in the sample statistics.

It is clear that the frequency of limit cycles decreases with the system size $N$. 
\begin{table*}[]
\begin{tabular}{|c|c|*{6}{r|}}
\hline
$N_1$&$N_{-1}$&$c_1$&$c_2$&$c_3$&$c_4$&$c_6$&$c_{12}$\\\hline
0&36&1&&&&&\\
1&35&36&&&&&\\
2&34&630&&&&&\\
3&33&7140&&&&&\\
4&32&58275&630&&&&\\
5&31&351792&25200&&&&\\
6&30&1713432&234360&&&&\\
7&29&7229700&1105020&12960&&&\\
8&28&23731605&6311655&194400&22680&&\\
9&27&68512360&23499000&1360800&362880&272160&136080\\
10&26&163379286&77290290&6259680&3084480&3265920&907200\\
11&25&405524196&164254860&20049120&3447360&3810240&3719520\\
12&24&798887565&365041215&49260960&14719320&20321280&3447360\\
13&23&1548952020&634738860&80831520&11793600&17781120&16692480\\
14&22&2467970100&1093051260&140019840&39009600&46811520&9434880\\
15&21&3798647280&1507217040&160885440&33203520&37195200&30754080\\
16&20&4770389610&2096265780&240226560&79606800&100154880&21228480\\
17&19&5850581940&2318217300&274337280&44089920&57062880&53207280\\
18&18&5862580080&2640716820&326894400&98703360&123379200&22861440\\
\hline
\multicolumn{2}{|c|}{$\sum$}&45674454016&19215221760&2273771520&557383680&696729600&301916160\\
\hline
\multicolumn{2}{|c|}{$\%$}&66.5&28.0&3.3&0.8&1.0&0.4\\
\hline
\end{tabular}
\caption{The numbers $c_k$ of initial states which lead to cycles of given length $k$ for $N=9$. In two first columns on the left, the number of positive $N_1$ and negative $N_{-1}$ links in the initial states are given. In the last row, the fraction of all lengths of the limit cycles are given, including those for $N_1>N_{-1}$, not shown here.}
\label{t2}
\end{table*}

\begin{table}[]
\begin{tabular}{ll}
$L_c$ & fraction  \\
\hline
1 & 0.8987022 \\
2 & 0.1007681 \\
3 & 0.0002326 \\
4 & 0.0001544 \\
6 & 0.0001416 \\
14 & 0.0000011 \\
\hline
\end{tabular}
\caption{The fractions of cycle length for $N=11$. The number of encountered cycles of length $14$ is $11172$  out of $10^{10}$ sampled configurations.}
\label{t5}
\end{table}

\section{Discussion}

Our result that the fraction $B$, of balanced states increases with $N$ is consistent with the results of \cite{11,12} obtained with the stochastic algorithm of Local Triad Dynamics. We note that the number of balanced states increases with $N$ as $2^{N-1}$, which is much slower than the number of all states: $2^{N(N-1)/2}$. The increase of $B$ with $N$ indicates that the common size of the basins of attraction of balanced states increases with the system size. By definition, basins of attraction of different states do not overlap.  Then we can evaluate an average size $S_B$ of the basin of attraction of a balanced state. As we read from Table \ref{N=9} and Fig. \ref{cc}, the probability of the balanced state at the end of evolution increases with $N$. This probability can be evaluated as $2^{N-1}S_B/2^{N(N-1)/2}$. This means that $S_B$ increases with $N$  roughly as $2^{N^2/2}$. Some evaluations \cite{24} indicate that all balanced states are equally probable. If this is the case, $S_B$ is just the size of the basin of attraction of each balanced state, and not only the average.

The number  $2^{N-1}$ of balanced configurations is just the number of Gribov copies of the configuration having all balanced triads. But as we discussed in the section on symmetries in fact any triad configuration has $2^{N-1}$ copies. The question is whether one can invent an algorithm which would avoid this $2^{N-1}$-fold degeneracy of triad configurations. The answer is affirmative.  One can use a gauge fixing procedure that exploits the invariance (\ref{gauge}) to fix signs of $(N-1)$ edges on a spanning tree.  Given a spanning tree and a configuration of signs $\{ s_{ij}\}$, one can namely use the invariance (\ref{gauge}) to choose $\sigma_i$'s to set $s'_{ij} = \sigma_i s_{ij} \sigma_j=1$ for every edge of the spanning tree. One can do this iteratively, edge by edge of the tree.  First by choosing $\sigma$'s at the endpoints of the first edge to set the edge sign to one, then by choosing $\sigma$ at the remaining endpoint of a neighboring edge, and repeating it for the next edge incident with either of the first two, etc. One can easily convince oneself that the freedom in choosing $\sigma_i$'s allows to do fix signs $s'_{ij}=1$  on any acyclic subgraph.  The spanning tree is just an acyclic connected subgraph containing all $N$ vertices. It has $(N-1)$ edges.   As a spanning tree we can choose a vertex with all $(N-1)$ edges attached to it and fix the signs of all these edges to one. The remaining edges form a complete graph $K_{N-1}$. The idea is now to update only edges on this $K_{N-1}$ subgraph using the original update rule (\ref{e1}), and leave the signs of the $(N-1)$ edges on the spanning tree.  In this way we reduce the problem from $K_N$ to $K_{N-1}$. The price to pay is that this update scheme breaks the symmetry of the original graph, while the emergence of cycles that we discussed seems to be deeply rooted in this symmetry,

Let us now argue why the probability of reaching the fully balance state asymptotically tends to unity for $N\rightarrow \infty$ for synchronous dynamics (\ref {e1}). We shall do this by comparing the synchronous  dynamics to the asynchronous one, that we shall try to understand first. As discussed, energy never increases when updates (\ref{e2}) are done asynchronously. It remains to understand what ensures that the process of energy lowering will not stop before the energy minimum is reached.  It might stop if there were no links in the system for which $U_{ij} > 0$ (\ref{e2}). One can show, however, that there is at least one link $ij$ such that $U_{ij} > 0$, as long as energy $U$ is greater than the minimal one. Thus the asynchronous algorithm will not stop till the energy minimum, and thus the full balance, is reached. For the synchronous updates (\ref{e1}) energy may increase. Let us suppose that $U_{ij}(t)>0$ and $U_{kl}(t)>0$ for two disjoint edges $ij$ and $kl$, and  $U_{ab}(t) \le 0$ for all other edges $ab$. Then the net change of energy is $U(t+1) = U(t) - 2 U_{ij}(t) - 2 U_{kl}(t)$, which is as if one updated edges asynchronously one after another, Eq. \ref{UUU}.  The situation gets complicated however when $U_{ij}(t)>0$ and $U_{jk}(t)>0$ for two incident edges $ij$ and $jk$, and  $U_{ab}(t) \le 0$ for all others $ab$. In this case  $U(t+1) = U(t) - 2 U_{ij}(t) - 2 U_{jk}(t) + 2 \Delta_{ijk}(t)$, because the triad $\Delta_{ijk}(t)$ is common for sets of triads the edges $ij$ and $jk$ belong to.  We see that the result of acting synchronously on edges having overlapping triads may increase energy as compared to the asynchronous update. In a more general situation, when there are more than two edges for which $U_{ij}>0$, energy of the system may increase $U(t+1)>U(t)$ after the update. This effect is frequently seen for small systems and it disappears when the systems size gets bigger. When the system size increases, the system gets diluted in the sense that  two incident edges have one common triad while each of them belongs to $N-2$ triads. Thus the interference from the overlap is $1/(N-2)$ compared to the asynchronous update and disappears for $N\rightarrow \infty$. Roughly speaking, for larger systems one can expect that the synchronous update scheme behaves like the asynchronous one for majority of configurations. For some configurations, however, due to their symmetry, limit cycles are observed.  They are sort of symmetry traps.  The simplest example is a configuration being a generalization of the configuration $C$ that we discussed for $N=5$. More generally, for $N=2n+1$, such a configuration consists of all positive links except  $n$ negative links which are incident with a vertex.  One can easily see that the synchronous update (\ref{e1}) will map this configuration into a configuration which also has only $n$ negative links incident with the vertex. They are complementary to those which were negative in the original configuration.  The two configurations form a limit cycle. We can give more explicit examples, however for the moment we have not a straightforward way of identifying configurations which lead to most of observed limit cycles, especially those of length $12$ or $14$. We find it a very challenging problem.

The difference between the results for even and odd values of $N$, shown in Table \ref{N=9}, can be attributed to the fact that for even $N$ the function sign(.) in the Eq. \ref{e1} can return zero. According to the update rule (\ref{e1}) in such a case the value of the link in question remains unchanged. As a consequence, the evolution towards the balance is stopped for some even values of $N$; hence $B$ is smaller and $J$ is larger there. However, for larger values of $N$ the zero value of the r.h.s. of Eq. \ref{e1} is less likely, and therefore the effect is weaker.

As shown in Section 5, the length $L_c$ of limit cycles increases around $N=9$, what makes this particular size specific. To interpret the effect, let us recall that $N=9$ has been identified in \cite{11,12} as the minimal size of the network where jammed states exist. In other words, $N=9$ is a boundary value between two regimes, where jammed states appear and do not appear. It is interesting that this boundary is visible in two different settings: the stochastic evolution towards the balance in \cite{11,12}, and the deterministic evolution modeled by Eq. \ref{e1}.  Perhaps long limit cycles can be interpreted as markers of shallow minima of a work function also in other deterministic models. 

The global coupling of links in our approach makes it similar to the N-K Kauffman model of genetic systems, with the highly correlated case $K=N-1$ \cite{25}.  In particular, the number of local optima is large (here $2^{N-1}$), and the length of adaptive walks to optima increases as logarithmic function of $N$. However, which evolves here is not the nodes but the links, which means that the counterpart of $N$ in the Kauffman model is the number of links, which is $L\approx N^2/2$. The evolution of a given link is determined by the state of $2(N-2)\approx \sqrt{8L}$ other links, which increases slower than the dimensionality $L$. This indicates that the similarity with the case $K=N-1$ is incomplete. 

Two comments may be added, one from a sociological perspective, and one from a computational point of view.  Our results, obtained within the cellular automata formalism, show that the process of removal cognitive dissonance in small groups might be counterproductive. In particular the number of imbalanced triads can be permanently larger than the number of balanced ones.  As the real process appears in conditions of imperfect information and dynamically varying situation \cite{26,27,28,29}, the importance of time sequence cannot be neglected. Yet this effect seems to loose its relevance for larger groups.

From the computational or dynamic point of view, the results indicate that the system of the size close to $N=9$ exhibits specific properties, namely one can relatively frequently observe non-trivial, long limit cycles. This seems to be the result of a compromise between two effects. On the one hand the length of limit cycles increases with the system size, on the other hand the frequency of cycles, relative to balanced states, decreases with the system size. The two effects meet for $N$ close to nine, in the sense that for systems  of the size in this range there are already longer cycles and on the other hand they are sufficiently abundant to be observed when states are sampled randomly.  We are not aware of any report on this effect in the literature.  We note that the rules on friends and enemies, mentioned in the introduction, are equivalent to negative XOR in Boolean Logic. Perhaps similar results could be obtained also for other Boolean functions.  In such a case systems of characteristic size should find application as loosely connected components of more complex entities. \\[1cm]
\noindent
{\Large \textbf{Acknowledgments}}\\[.3cm]
We are grateful to Rafa{\l} Kalinowski for helpful comments. This research was supported in part by PLGrid Infrastructure.


\begin{thebibliography}{30}

\bibitem{1} C. Castellano, S. Fortunato and V. Loreto, Statistical physics of social dynamics, Rev. Mod. Phys. 81, 591-646 (2009)

\bibitem{2} P. Sen and B. K. Chakrabarti, Sociophysics. An Introduction, Oxford UP, 2013

\bibitem{3} F. Schweitzer, Sociophysics, Physics Today 71, 2, 40 (2018)

\bibitem{4} S. Galam, Sociophysics. A Physicist's Modeling of Psycho-political Phenomena, Springer-Verlag 2012.

\bibitem{5} S. Thurner, R. Hanel and P. Klimek, Introduction to the Theory of Complex Systems, Oxford UP 2018.  

\bibitem{6} S. P. Borgatti, M. G. Everett and J. C. Johnson, Analyzing Social Networks, Sage Publ., LA 2013.

\bibitem{7} L. Festinger, A Theory of Cognitive Dissonance, Stanford University Press, Stanford, 1957.

\bibitem{8} F. Heider,  Attitudes  and  cognitive  organization,  The  Journal of Psychology 21 (1) (1946) 107-112.

\bibitem{9} D. Cartwright and F. Harary, Structural  balance:   A  generalization of Heider's theory, Psychological 
Review 63 (1956) 277-293.

\bibitem{10} P. Bonacich and P. Lu, Introduction to Mathematical Sociology, Princeton UP, 2012

\bibitem{11} T. Antal, P. L. Krapivsky and S. Redner, Dynamics of social balance on networks, Physical Review E 72 (2005) 036121.

\bibitem{12} T. Antal, P. Krapivsky and S. Redner, Social balance on networks: The  dynamics  of  friendship  and  enmity, 
 Physica  D: 224 (1) (2006) 130-136.

\bibitem{13} K.  Ku{\l}akowski,  P.  Gawro\'nski and  P.  Gronek,  The  Heider  balance: A continuous approach,    
International Journal of Modern  Physics  C  16  (5)  (2005)  707-716.

\bibitem{14} S.  A.  Marvel,  J.  Kleinberg,  R.  D.  Kleinberg and  S.  H.  Strogatz, Continuous-time  model  of  structural  balance,  
Proceedings  of the  National  Academy  of  Sciences  108  (5)  (2011)  1771-1776.
 
\bibitem{15} K. Malarz, M. Wo{\l}oszyn and K. Ku{\l}akowski, Towards the Heider balance with a cellular automaton, Physica D 411 (2020) 132506. 

\bibitem{16} K. Malarz and K. Ku{\l}akowski, Heider balance of a chain of actors as dependent on the interaction range and a thermal noise,
Physica A 567 (2021) 125689.

\bibitem{17} F.  Rabbani,  A.  H.  Shirazi,   and  G.  R.  Jafari,  Mean-field solution of structural balance dynamics in nonzero 
temperature, Physical Review E99, 062302 (2019)
 
\bibitem{18} R.  Shojaei,  P.  Manshour,   and  A.  Montakhab,  Phase transition  in  a  network  model  of  social  balance  with 
Glauber  dynamics,  Physical  Review  E 100,  022303 (2019).

\bibitem{19} S. Arabzadeh, M. Sherafati, F. Atyabi, G.R. Jafari and K. Ku{\l}akowski, Lifetime of links influences the evolution towards 
structural balance, Physica A 567 (2021) 125689.

\bibitem{20} F. Hassanibesheli, L. Hedayatifar, P. Gawro\'nski, M. Stojkow, D. \.Zuchowska-Skiba and K. Ku{\l}akowski, Gain and loss of esteem, 
direct reciprocity and Heider balance, Physica A 468 (2017) 334.

\bibitem{21} M. J. Krawczyk, M. Wo{\l}oszyn, P. Gronek, K.Ku{\l}akowski and J. Mucha, The Heider balance and the looking-glass self: modelling 
dynamics of social relations, Sci. Rep. 9 (2019) 11202. 

\bibitem{str} P. Straffin, Game Theory and Strategy, Math. Assoc. of America, Washington D.C., 1993.

\bibitem{22} E. Aronson, V. Cope, My enemy's enemy is my friend, Journal of  Personality  and  Social  Psychology  8  (1968)  8-12.

\bibitem{23} K. Klemm and S. Bornholdt, Stable and unstable attractors in Boolean networks, Phys Rev E 72 (2005) 055101.

\bibitem{gri} V. N. Gribov, Quantization of non-Abelian gauge theories, Nucl. Phys. B139 (1978) 1 

\bibitem{24} M. J. Krawczyk, S. Ka{\l}u\.zny and K. Ku{\l}akowski, A small chance of paradise - equivalence of balanced states, EPL 118 (2017) 58005.

\bibitem{25} S. A. Kauffman, The Origins of Order. Self-Organization and Selection in Evolution, Oxford University Press, New York 1993.

\bibitem{26} J. W. Brehm, Increasing cognitive dissonance by a fait accompli, Journal of Abnormal and Social Psychology 58 (1959) 379-382.

\bibitem{27} J. D. Jecker, Conflict and dissonance: a time of decision, in Theories of Cognitive Consistency: A Sourcebook, ed. by R. P. Abelson,
 E. Aronson, W. J. McGuire, Th. M. Newcomb, M. J. Rosenberg and P. H. Tannenbaum, Rand McNally and Co., Chicago 1968.
 
\bibitem{28} J. E. Russo, M. G. Meloy and V. H. Medvec, Predecisional distortion of product information,  J. of Marketing Research 35 (1998) 438-452. 

\bibitem{29} Y. H. Liang, Reading to make a decision or to reduce cognitive dissonance? The effect of selecting and reading online reviews 
from a post-decision context,  Computers in Human Behavior 64 (2016) 463-471.

\end{thebibliography}

\end{document}